# Multilayered feed forward Artificial Neural Network model to predict the average summer-monsoon rainfall in India

Surajit Chattopadhyay

Department of Mathematics, Techno Model School

Techno India Group

EM-4/1, Sector V, Salt Lake

Kolkata 700 091

India

Ph +91339830736116

E-mail surajit_2008@yahoo.co.in

## Abstract

In the present research, possibility of predicting average summer-monsoon rainfall over India has been analyzed through Artificial Neural Network models. In formulating the Artificial Neural Network based predictive model, three layered networks have been constructed with sigmoid non-linearity. The models under study are different in the number of hidden neurons. After a thorough training and test procedure, neural net with three nodes in the hidden layer is found to be the best predictive model.

# 1. Introduction

Weather forecasting is one of the most imperative and demanding operational responsibilities carried out by meteorological services all over the world. It is a complicated procedure that includes numerous specialized fields of know-how (Guhathakurata, 2006). The task is complicated because in the field of meteorology all decisions are to be taken in the visage of uncertainty. Several authors (i.e. Brown and Murphy, 1988; Handerson and Wells, 1988; Wilks, 1991; Elsner and Tsonis, 1992 and many others) have discussed the uncertainty associated with the weather systems. Chaotic features associated with the atmospheric phenomena also have attracted the attention of the modern scientists (Sivakumar 2000, 2001; Sivakumar et al, 1999; Men et al, 2004). Different scientists over the globe have developed stochastic weather models which are basically statistical models that can be used as random number generators whose output resembles the weather data to which they have been fit (Wilks, 1999). Statistical models have the drawback that in most of the cases they depend upon several tacit assumptions regarding the system. But, a chaotic system like atmosphere cannot be bound by any postulation. The numerical models are based on the scheme of nonlinear operator equations prevailing the atmospheric system. But in the absence of any analogue solution of this system of operator equations, numerical solutions based on different assumptions are the only alternative. Furthermore, the chaotic behaviors of these nonlinear equations sensitive to initial conditions make it more intricate to solve these equations (Guhathakurta, 2006). As a result, flawed forecast comes out.

Amongst all weather happenings, rainfall plays the most imperative role in human life. Human civilization to a great extent depends upon its frequency and amount to various scales. Several stochastic models have been attempted to forecast the occurrence of rainfall, to investigate its seasonal variability, to forecast monthly/ yearly rainfall over some given geographical area. Daily precipitation occurrence has been viewed through Markov chain by (Chin, 1977). Gregory et al (1993) applied a chain-dependent stochastic model, named as Markov chain model to investigate inter annual variability of area average total precipitation. Wilks (1998) applied mixed exponential distribution to simulate precipitation amount at multiple sites exhibiting realistic spatial correlation.

Present paper endeavors to develop an Artificial Neural Network (ANN) model to forecast average rainfall during summer-monsoon in India. The term "monsoon" seems to have been derived either from the Arabic *mausin* or from the Malayan *monsin*. As first used it was applied to southern Asia and the adjacent waters, where it referred to the seasonal surface air streams which reverse their directions between winter and summer, southwest in summer and northeast in winter in this area. In 1686 Halley explained the Asiatic monsoon as resulting from thermal contrasts between the continent and oceans. During the summer the continent is heated, leading to rising motion and lower pressure. This induces airflow from sea to land at low elevations.

Eastern and southern Asia has the earth's largest and best-developed monsoon circulations. The tropical monsoon circulation of southern Asia, including India-Pakistan and Southeast Asia, differs significantly from East Asia monsoon. The Indian monsoon is effectively separated from that of China by the Himalayan-Tibet system. In summer a deep and widespread surface pressure trough extends across northern India-Pakistan into Southeast Asia. This is part of the planetary intertropical convergence zone, which here reaches its maximum pole ward displacement. To the south of the trough is a deep current of maritime tropical air called the southwest monsoon. This current appears to originate in the southeast trades of the Indian Ocean east of Africa. As this stream of air approaches and crosses the equator its direction becomes southerly and then southwesterly. Along the Somali coast of Africa the flow becomes especially strong, taking the form of a low-level jet. In crossing the Arabian Sea the southwesterly current gains considerable moisture and becomes less stable. This unstable southwesterly current crosses India, continues eastward over the Indochina peninsula, and then move northward over much of eastern Asia. It is a great moisture source for most of southern Asia.

Indian economy is standing on Indian summer monsoon. So prediction of Indian summer monsoon is a challenging topic to Indian atmospheric scientists. Hastenrath (1988) developed statistical model using regression method to predict Indian summer monsoon rainfall anomaly. Rajeevan (2001) discussed the problems and prospects in prediction of Indian summer monsoon and revealed that Indian summer monsoon predictability exhibits epochal variations. Gadgil et al (2005) investigated the causes of failure in prediction of Indian summer monsoon and expected the dynamical models to generate better prediction only after the problem of simulating year-to-year variation of monsoon is addressed. Kishtawal et al (2003) assessed the feasibility of a nonlinear technique based on genetic algorithm, an Artificial Intelligence technique for the prediction of summer rainfall over India. Guhathakurta (2006) implemented ANN technique to predict rainfall over a state (Kerala) of India, and to the best of the knowledge and belief of the authors of the present study, Guhathakurta (2006) is the first ever attempt to implement

ANN to forecast the summer-monsoon over India. But, Guhathakurta (2006) confined his study within a state of India. Present contribution deviates from the study of Guhathakurta (2006) in the sense that instead of choosing a particular state, the authors implement Backpropagation ANN to forecast the average summer-monsoon rainfall over the whole country and aroma of newness further lies in the fact that here various multilayer ANN models are attempted to find out the best fit.

## 2. ANN in rainfall prediction – a literature survey

Hu (1964) initiated the implementation of ANN, an important Soft Computing methodology in weather forecasting. Since the last few decades, ANN a voluminous development in the application field of ANN has opened up new avenues to the forecasting task involving atmosphere related phenomena (Gardner and Dorling, 1998; Hsieh and Tang, 1998). Michaelides et al (1995) compared the performance of ANN with multiple linear regressions in estimating missing rainfall data over Cyprus. Kalogirou et al (1997) implemented ANN to reconstruct the rainfall time series over Cyprus. Lee et al (1998) applied Artificial Neural Network in rainfall prediction by splitting the available data into homogeneous subpopulations. Wong et al (1999) constructed fuzzy rule bases with the aid of SOM and Backpropagation neural networks and then with the help of the rule base developed predictive model for rainfall over Switzerland using spatial interpolation.

## 3. Materials and method

ANNs have recently become important alternative tool to conventional methods in modelling complex non-linear relationships. In the recent past, the ANN has been applied to model large data with large dimensionality (i.e. Gevrey et al., 2003; Nagendra and Khare, 2006). Most of the ANN studies spoke to the problem allied with pattern recognition, forecasting and comparison of the neural network with other traditional approaches in ecological and atmospheric sciences. However, the step-by-step procedure involved in development of ANN-based models is less discussed (Nagendra and Khare, 2006). This paper develops ANN model step-by-step to predict the average rainfall over India during summer- monsoon by exploring the data available at the website [http://www.tropmet.res.in](http://www.tropmet.res.in) published by Indian Institute of Tropical Meteorology.

The ANN approach has several advantages over conventional phenomenological or semi-empirical models, since they require known input data set without any assumptions (Gardner and Dorling, 1998; Nagendra and Khare, 2006). It exhibits rapid information processing and is able to develop a mapping of the input and output variables. Such a mapping can subsequently be used to predict desired outputs as a function of suitable inputs (Nagendra and Khare, 2006). A multilayer neural network can approximate any smooth, measurable function between input and output vectors by selecting a suitable set of connecting weights and transfer functions or activation function (Gardner and Dorling, 1998; Kartalopoulos, 1996; Nagendra and Khare, 2006).

## 3.1  ANN based prediction of summer-monsoon rainfall in India

The model building process consists of four sequential steps:

(i) Selection of the input and output for the supervised Backpropagation learning

(ii) Selection of the activation function

(iii) Training and testing of the model

(iv) Testing the goodness of fit of the model

### 3.1.1 Selection of the input and output for the supervised Backpropagation learning

In India, the months June, July, and August are identified as the summer-monsoon months. Thus, the present study explores the data of these three months corresponding to the years 1871-1999. From these 129 years, last year is deleted because that would not lead to any prediction. Thus, there would be (128×3=384) months in our modeling problem. For each month, there would be a time series of homogenized rainfall data with 128 entries. It is interesting to see that the time series are not pair wise correlated. The mutual Pearson correlation values are –0.06 (June-July), -0.01 (June-August), and –0.01(July-August). Thus, all the correlation values are too small, indicating that the relationships are highly non-linear. Thus, necessity of implementing ANN in the prediction problem is felt highly relevant. Furthermore, the autocorrelations for each month are found to be significantly small (Fig.01). This indicates that the data exhibit no serial correlation or persistence. Aim of this paper is to develop a multilayer feed forward ANN model so that the average summer-monsoon rainfall of a given year can be predicted using the rainfall data of the summer-monsoon months of the immediately previous year. Thus, the input matrix would consist of four columns of which the first three columns would correspond to the summer-monsoon months' rainfall of year '$n$' and the fourth column would correspond to the average summer-monsoon rainfall of the year ($n+1$). Basically, the fourth column would correspond to the 'desired output' in the supervised Backpropagation learning (Kartalopoulos, 1996) procedure. The first 75% data (i.e. 96 rows out of 128 rows) are taken as the training set and the remaining 25% data (i.e. 32 rows out of 128 rows) are taken as the test set or validation set.

To avoid the asymptotic effect the raw data are scaled according to

$$z_i = 0.1 + 0.8 \times \left( \frac{x_i - x_{\min}}{x_{\max} - x_{\min}} \right) \quad \ldots \quad \ldots \quad \ldots \quad (1)$$

Where, $z_i$ denotes the transformed appearance of the raw data $x_i$.

After the modeling is completed, the scaled data are reverse scaled according to

$$P_i = x_{\min} + \left( \frac{1}{0.8} \right) \times \left[ (y_i - 0.2) \times (x_{\max} - x_{\min}) \right] \quad \ldots \quad \ldots \quad (2)$$

Where, $P_i$ denotes the prediction in original scale, and the corresponding scaled prediction is $y_i$.

### *3.1.2 Activation function*

The advent of Backpropagation algorithm (BP), the adaptation of steepest descent method, opened up new avenues of application of Multilayered ANN for many problems of practical interest (see e.g. Perez and Reyes, 2001; Kamarthi and Pittner, 1999; Sejnowski and Rosenberg, 1987). A multilayer ANN contains three basic types of layer: input layer, hidden layer (s), and output layer. Basically the Backpropagation learning involves propagation of error backwards from the output layers to the hidden layers in order to determine the update for the weights leading to the units in the hidden layer(s). The methodology is detailed in section 3.1.3.

The non-linear relationship between input and output parameters in any network requires a function, which can appropriately connect and/or relate the corresponding parameters (Nagendra and Khare, 2006). In the present paper Backpropagation learning of ANN would be adopted with steepest descent (Kartalopoulos, 1996). Thus, the sigmoidal function is taken as the ideal activation mathematically defined as (Kartalopoulos, 1996)

$$f(z) = \frac{1}{1+\exp(-z)} \qquad \ldots \qquad \ldots \qquad \ldots \qquad (3)$$

### *3.1.3 Training and testing of the model*

The proposed ANN model is basically a three layered ANN with Backpropagation learning. Before going into the implementation details, the Backpropagation algorithm for Multilayered ANN is briefly discussed.

Backpropagation learning is exactly the same as delta learning at the output layer and is similar to the delta learning with the propagated error in the hidden layer(s), and thus it is called generalized delta rule (Yegnanarayana, 2000). In this algorithm, at first step, the input and desired output are identified. Then an arbitrary weight vector $w_0$ is initialized. Then the feed forward neural network is iteratively adopted according to the recursion (Kamarthi and Pittner, 1999)

$$w_{k+1} = w_k + \eta d_k \qquad \ldots \qquad \ldots \qquad \ldots \qquad (4)$$

Where, $w_l$ denotes the weight matrix at epoch $l$. the positive constant $\eta$, which is selected by the user, is called the learning rate.

The direction vector $d_k$ is negative of the gradient of the output error function $E$

$$d_k = -\nabla E(w_k) \qquad \ldots \qquad \ldots \qquad \ldots \qquad (5)$$

There are two standard learning schemes for the BP algorithm: on-line learning and batch learning. In on-line learning, the weights of the network are updated immediately after the presentation of each pair of input and target patterns. In batch learning all the pairs of patterns in the training sets are treated as a batch, and the network is updated after processing of all training patterns in the batch. In either case the vector $w_k$ contains the weights computed during $k$th iteration, and the output error function $E$ is a multivariate function of the weights in the network (Kamarthi and Pittner, 1999)

$$E(w_k) = \begin{cases} E_p(w_k) [on-line] \\ \sum_p E_p(w_k) [Batch] \end{cases} \quad \ldots \quad \ldots \quad \ldots \quad (6)$$

Where, $E_p(w_k)$ denotes the half-sum-of-squares error functions of the network output for a certain input pattern $p$. The purpose of the supervised learning (or training) is to find out a set of weight that can minimize the error $E$ over the complete set of training pair. Every cycle in which each one of the training patterns is presented once to the neural network is called an epoch.

The direction vector $d_k$, expressed in terms of error gradient depends upon the choice of activation function. When the sigmoid function (as defining equation (3)) is adopted, the BP algorithm becomes 'Back propagation for the Sigmoid Adaline' (Widrow and Lehr, 1990). In this method the input matrix is multiplied by the weight matrix and the product is used as the variable for the sigmoid activation function. For example, at epoch $k$, the sigmoid non linearity is produced as

$$f(W_k X_k) = \frac{1}{1 + e^{-\left(\sum_i w_i x_i\right)}} \quad \ldots \quad \ldots \quad \ldots \quad (7)$$

Where $W_k = [w_1 \quad w_2 \quad \ldots \quad w_n]$ and $X_k = [x_1 \quad x_2 \quad \ldots \quad x_n]^T$ are the weight matrix and the transpose of the input matrix respectively at epoch $k$.

After training or learning the ANN with BP algorithm with sigmoid non-linearity, a ultimate weight matrix is obtained. This weight matrix is applied to another set of independent inputs to examine the efficiency of the model. This phase is called the testing or the validation phase.

### *3.1.4 Testing the goodness of fit of the model*

After developing the model through training and testing, goodness of fit of the model is examined statistically. Over all prediction error (PE) is measured as (Perez and Reyes, 2001)

$$PE = \frac{\langle |y_{predicted} - y_{actual}| \rangle}{\langle y_{actual} \rangle} \quad \ldots \quad \ldots \quad \ldots \quad (8)$$

Where, $\langle \; \rangle$ implies the average over the whole test set.

The predictive model is identified as a good one if the *PE* is sufficiently small i.e. close to 0. The model with minimum *PE* is identified as the best prediction model.

## 4. Implementation details and the results

Details of the input and output variables are presented in section 3.1.1. The learning rate $\eta$ (see equation (4)) is taken to be 0.9. A three-layered feed forward neural net is now designed. The problem is to find out the number of hidden nodes producing the best model. Since the number of adjustable parameters in a one hidden-layer feed forward neural network with $n_i$ input units, $n_0$ output units, and $n_h$ hidden units is $[n_0 + n_h(n_i + n_0 + 1)]$ (Perez et al, 2000) for $n_i = 3$, $n_0 = 1$ and with 96 training cases, it is not possible to use an $n_h$ greater than 19.

Now, 19 three-layered feed forward ANN models with $n_h = 1,2,3,\ldots\ldots,19$ and $\eta = 0.9$ are generated. Model $M_k$ would imply the three-layered feed forward ANN with $k$ nodes in the hidden layered and trained through on-line (ref equation (6)) Backpropagation learning using the methodology explained in section 3.1.3. In all the 19 models the initial weights are chosen randomly from –0.5 to +0.5 (Pal and Mitra, 1999). After each training iterations/epochs the network is tested for its performance on validation data set. The training process is stopped when the performance reach the maximum on validation data set (Haykin, 2001; Sarle, 1997;Gardner and Dorling, 1998; Nagendra and Khare, 2006). After training and testing, the *PE* (ref equation (8)) values are computed for each model. The results are schematically presented in Fig.02.

The result shows that the model $M_3$ produces the lowest prediction error among the 19 possible predictive models. After 500 epochs the final weight matrix for $M_3$ is found to be

$$\begin{array}{c} Hidden\_Nrn1 \\ Hidden\_Nrn2 \\ Hidden\_Nrn3 \end{array} \begin{bmatrix} 0.1902 & -0.5335 & -0.3964 & -0.3215 \\ -1.0064 & 1.1778 & 2.6772 & 1.6823 \\ -4.0503 & 1.4562 & 2.0527 & 1.9305 \end{bmatrix}_{3\times 4}$$

In Fig.03, the performance of $M_3$ is pictorially presented. This figure shows that, the actual average summer-monsoon rainfall in India has a close association with those predicted through Multilayered feed forward ANN (sigmoid non-linearity) with 3 nodes in the hidden layer. In the second y-axis, the prediction errors (%) in each test case are presented. It is evident from the figure that in 78.12% cases, the prediction error is below 30%, and in 50% cases the prediction error is below 20%. These results, undoubtedly, reflect high prediction yield from the $M_3$ model. The components of $M_3$ are presented in Table 01.

## 5.   Conclusion

After comparing the performance of 19 three layered ANN models with sigmoid non linearity, the ANN model with 3 hidden nodes is found to be adroit for prediction of mean summer-monsoon rainfall over India on the basis of previous years' rainfall data of the summer-monsoon months.

Table01-Basic network components of the $M_3$ model

## Network Architecture

| | | | |
|---|---|---|---|
| Number of Inputs | **3** | Number of Outputs | **1** |
| Number of Hidden Layers | **1** | | |
| | | Hidden Layer sizes | **3** |
| Learning parameter | **0.2** | Initial Wt Range (0 +/- w) | **0.5** |
| Momentum | **0.9** | | |
| Training Options | | | |
| Total #rows in your data | **128** | No of Training cycles | **500** |
| | | Training Mode | **On line** |
| Save Network weights | **With least Training Error** | | |
| Training / Validation Set | **Partition data into Training / Validation set** | | |

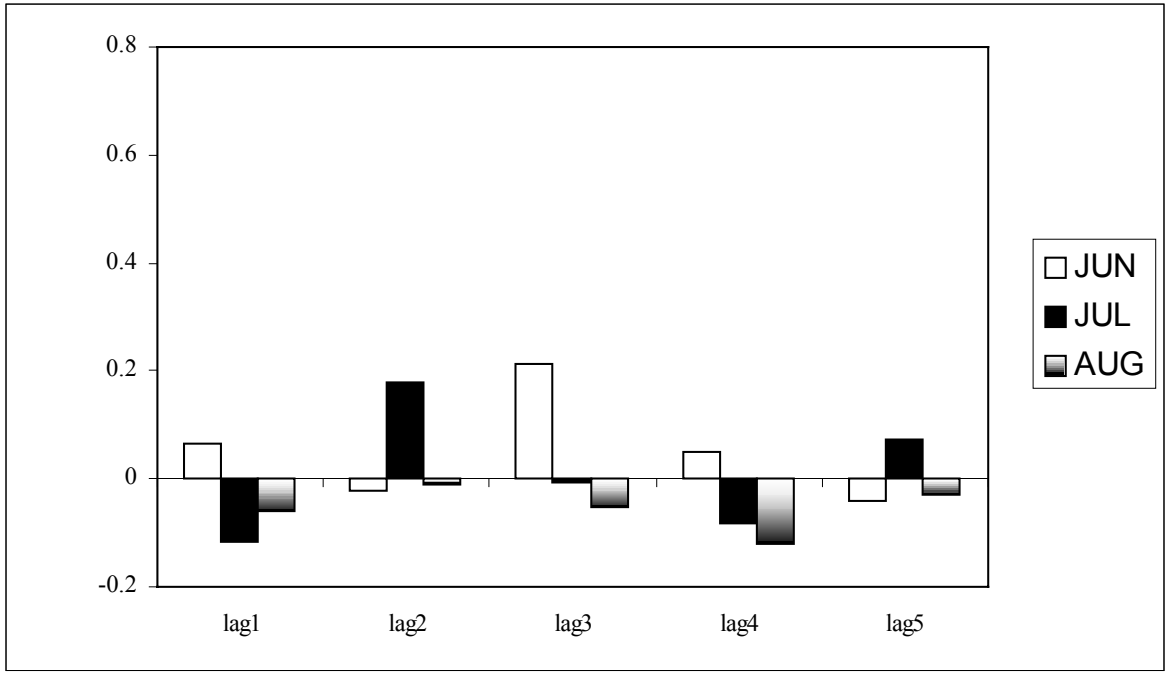

Fig.01- Schematic showing the autocorrelation function corresponding to the rainfall amounts in the summer monsoon months between 1871 and 1999.

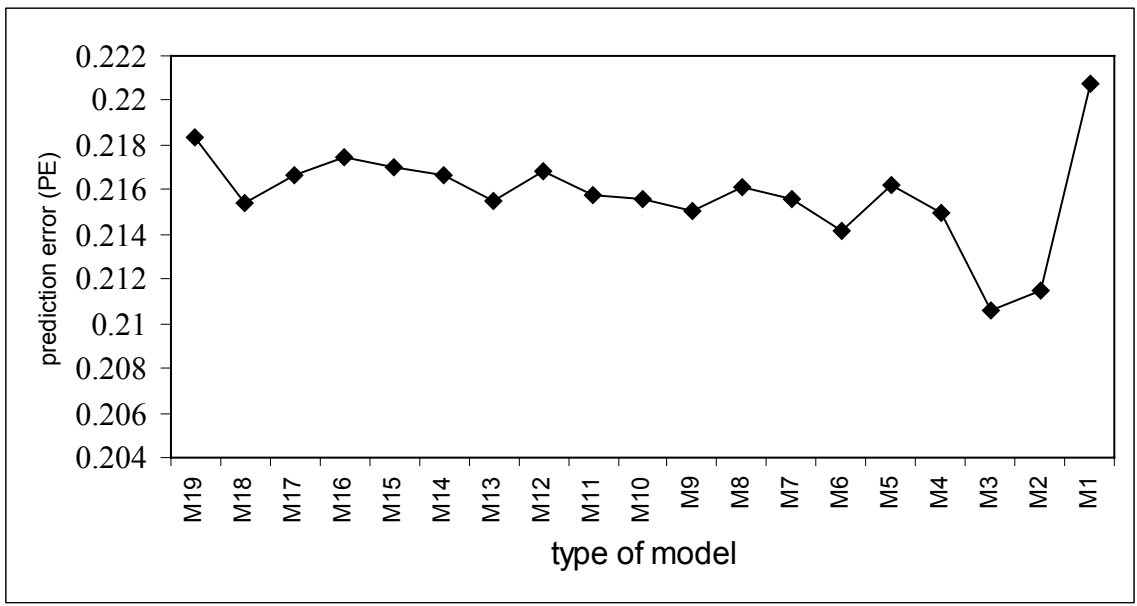

Fig.02- Schematic showing the prediction error over the whole test set corresponding to 19 different three layered ANN models.

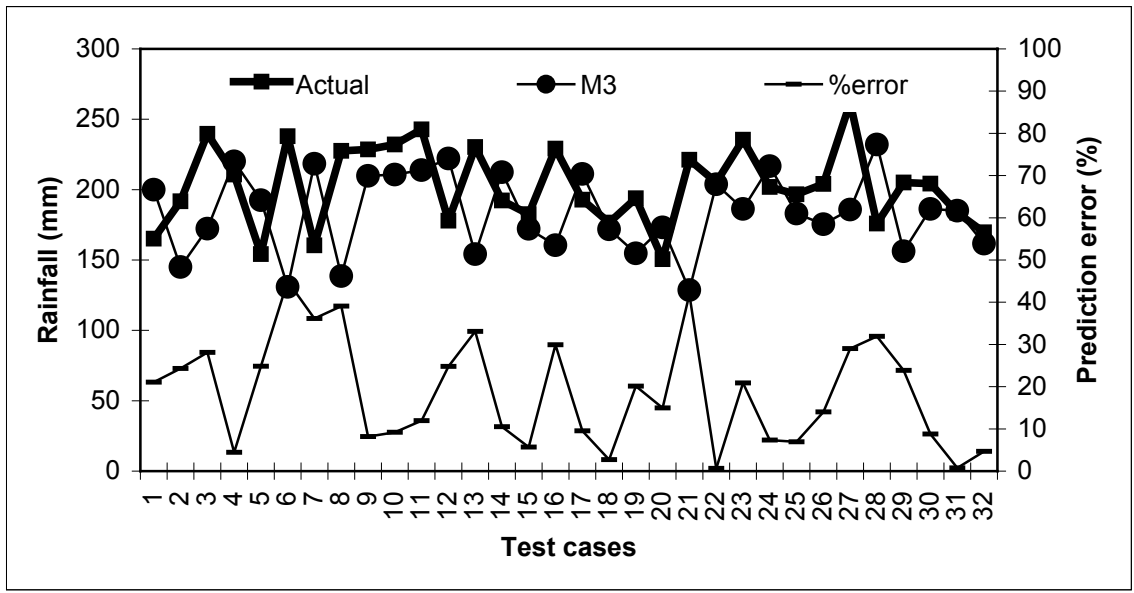

Fig.03- Schematic showing the actual and predicted (by $M^3$) average summer-monsoon rainfall in the test cases. In the second y-axis the percentage errors of prediction in all the test cases.